\documentstyle[aps,prl,epsfig]{revtex}
\begin{document}
\twocolumn[\hsize\textwidth\columnwidth\hsize\csname %
@twocolumnfalse\endcsname       
 
\title{Electron Momentum Distribution Function in the $t$-$t'$-$J$
Model}
\author{A. Ram\v sak$^{1,2}$, I. Sega$^1$, and  P. Prelov\v sek$^{1,2}$,}
\address{ $^{1}$ J. Stefan Institute, 1000 Ljubljana, Slovenia }
\address{$^{2}$ Faculty of Mathematics and Physics, University of
Ljubljana, 1000 Ljubljana, Slovenia }
\date{\today}
\maketitle
\begin{abstract}\widetext
We study the electron momentum distribution function (EMDF) for the
two-dimensional $t$-$t'$-$J$ model doped with one hole on finite
clusters by the method of twisted boundary conditions. The results
quantitatively agree with our analytical results for a single hole in
the antiferromagnetic background, based on the self-consistent Born
approximation (SCBA).  Moreover, within the SCBA an anomalous momentum
dependence of EMDF is found, pointing to an emerging large Fermi
surface. The analysis shows that the presence of next-nearest-neighbor
(NNN) hopping terms changes EMDF only quantitatively if the ground
state (GS) momentum is at $({\pi \over 2},{\pi \over 2})$ and
qualitatively if the GS momentum is shifted to $(\pi,0)$.
\end{abstract}
\pacs{PACS numbers: 71.27.+a, 72.15.-v} ]
\narrowtext

\section{Introduction}

The existence and the character of the Fermi surface (FS) of high
$T_c$ superconductors, in particular in their underdoped regime, is
one of the open questions \cite{shen95} of solid state physics.  The
key quantity for resolving this problem is the electron momentum
distribution function $n_{\bf k}=\langle
\Psi_{{\bf k}_0}|\sum_\sigma c_{{\bf k},\sigma}^\dagger c_{{\bf
k},\sigma} |\Psi_{{\bf k}_0} \rangle$.  Here we study the EMDF for
$|\Psi_{{\bf k}_0} \rangle$ which represents a weakly doped
antiferromagnet (AFM), i.e., it is the GS wave function of a planar
AFM with {\it one hole} and the GS wave vector ${\bf k}_0$. The wave
function is determined within the framework of the standard $t$-$J$
model with nearest-neighbor hopping $t_{jj'}\equiv t$ and the AFM
exchange is fixed to $J=0.3t$.  In order to come closer to the
realistic situation in cuprates the model is extended with the NNN
hopping $t_{jj'}\equiv t^\prime$ for $jj'$ representing
next-nearest-neighbors,
\begin{eqnarray}
H&=& -\sum_{<jj^\prime>\sigma} t_{jj^\prime} \bigl(
{\tilde c}_{j,\sigma}^\dagger
{\tilde c}_{j^\prime,\sigma} + \mbox{H.c.} \bigr)+ \nonumber \\
& &+ J \sum_{<ij>}
\bigl[ S_i^z S_j^z+\frac{\gamma}{2}(S_i^+ S_j^- + S_i^- S_j^+ ) \bigr].
\label{tj}
\end{eqnarray}
Numerically $n_{\bf k}$ can be determined by exact diagonalization
(ED) of the model Eq.~(\ref{tj}) in small clusters, where only a
restricted number of momenta ${\bf k}$ is allowed. The GS wave vector
due to finite size effects varies with $N$. Therefore we present here
results obtained with the method of twisted boundary conditions
\cite{zotos}, where $t_{jj'} \to t_{jj'} \exp{i {\theta}_{jj'}}$.
Since $n_{\bf k}\equiv n_{\bf k}({\bf k}_0,{\bf \theta})$ depends both
on ${\bf k}_0$ and ${\bf \theta}$ it follows from Peierls construction
that $n_{\bf k}({\bf k}_0,0)=n_{{\bf k}+{\bf k}_0}(0,{\bf k}_0)$ for
${\bf \theta}={\bf k}_0$. This allows us to study $n_{\bf k}$ for
arbitrary ${\bf k}$ and ${\bf k}_0$.  Furthermore, the finite size
effects of the results are suppressed if we fix ${\bf k}_0$ for {\it
all clusters} here studied to the symmetry point ${\bf
k}_0=(\frac\pi2,\frac\pi2)$.

\section{Results}

In Fig.~1(a) we present ED results for clusters with different $N$ and
$\gamma=1$. The EMDF obeys the sum rule $\sum_{\bf k} n_{\bf k}=N-1$
and, for the allowed momenta, the constraint $N (n_{\bf
k}-1)\leq1$. We show here the quantity $N (n_{\bf k}-1)$, which for
different $N$ scales towards the same curve.  Results are presented
for momenta ${\bf k}\in[(-\pi,-\pi) \to (\pi,\pi)]$ and should be
averaged over all four possible ground state momenta when discussed,
e.g., in connection with ARPES data.

We further compare the results with the self-consistent Born
approximation \cite{schmitt88}, where we decouple fermion operators
into hole and magnon operators
\cite{martinez91,ramsak93}, and using the SCBA wave function
\cite{ramsak93,ramsak99} we evaluate the 
required matrix elements in $n_{\bf k}$. In Fig.~1(a) the result for
$N (n_{\bf k}-1)$ is presented. In the SCBA there appear in $n_{\bf
k}$ at ${\bf k}\sim {\bf k}_0,{\bf k}_0+(\pi,\pi)$ also delta-function
contributions proportional to the quasiparticle pole residue, $Z_{{\bf
k}_0}$, but {\it not shown} in the figures.  The comparison of the
SCBA with the ED results shows a quantitative agreement, however, a
surprising observation is that $n_{\bf k}$ in the SCBA exhibits a
discontinuity for momenta ${\bf k}\sim {\bf k}_0,{\bf k}_0+(\pi,\pi)$
\cite{ramsak99}. 
We interpret this result as an indication of an emerging {\it large}
Fermi surface at ${\bf k}\sim \pm {\bf k}_0$ indicatingt the
coexistence of two apparently contradicting FS scenarios in EMDF of a
single hole in an AFM. On one hand, the $\delta$-function
contributions seem to indicate that at finite doping a delta-function
might develop into small FS (hole pockets). On the other hand the
discontinuities at ${\bf k}={\bf k}_0$, ${\bf k}_0+(\pi,\pi)$ are more
consistent with infinitesimally short arcs (two points) of an emerging
large FS.

In Fig.~1(b) and Fig.~1(c) we also present the results for the
$t$-$t^\prime$-$J$ model.  (i) The effect of {\it negative} NNN
$t'=-t/4$ is relatively weak: the GS momentum remains at ${\bf
k}_0=(\frac\pi2,\frac\pi2)$. Otherwise results are similar to the
$t'=0$ case.  (ii) {\it Positive} NNN hopping matrix elements
$t'=t/4$: the GS is now twofold degenerate, with the momenta at
corners of the AFM zone, e.g., ${\bf k}_0=(\pi,0)$. All results agree
with the SCBA method.

\newpage

\begin{figure}[b]         
\begin{center}\leavevmode  
\includegraphics[width=0.8\linewidth,angle=-90]{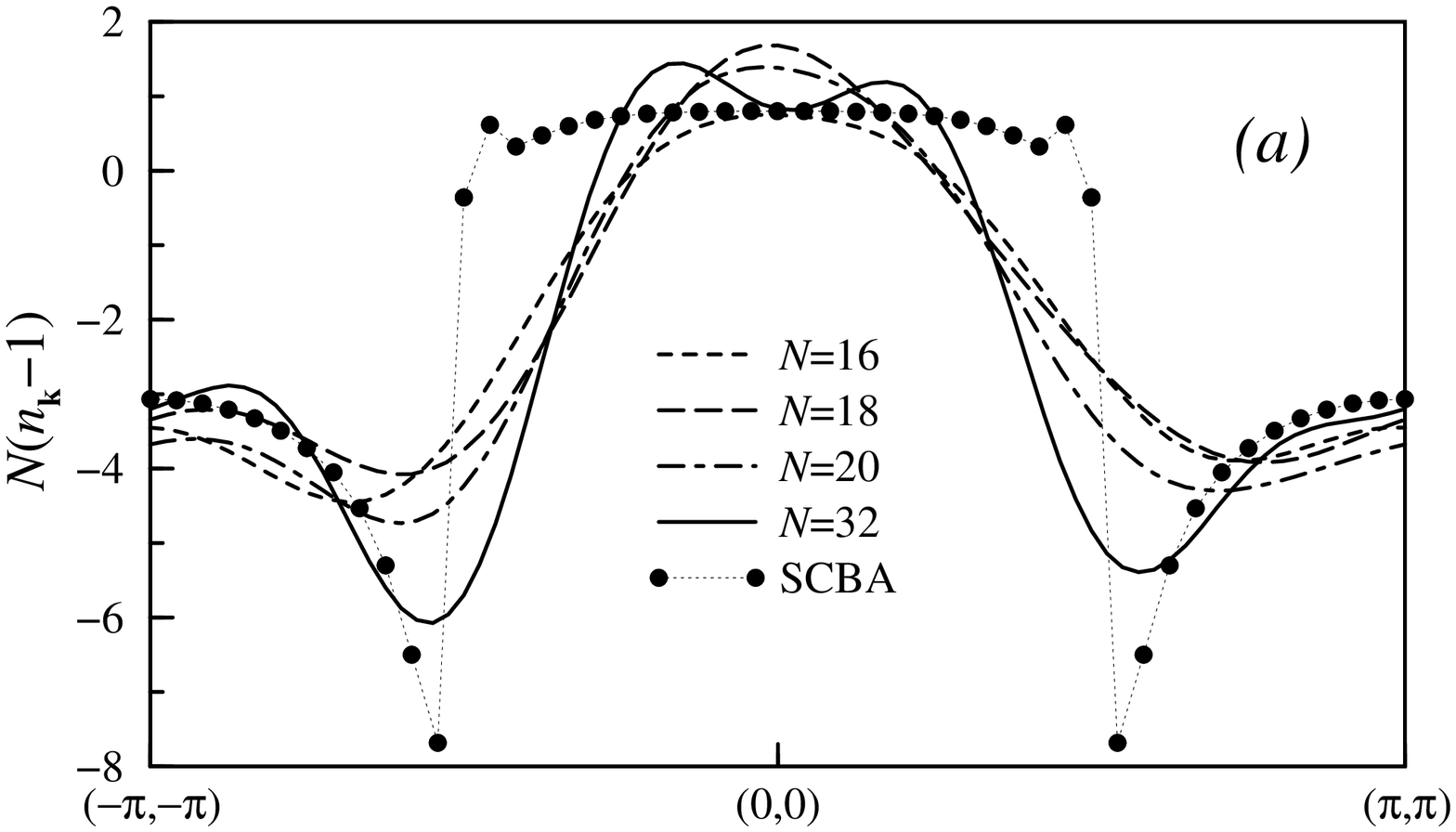}
\vskip -2.5  cm    
\includegraphics[width=0.8\linewidth,angle=-90]{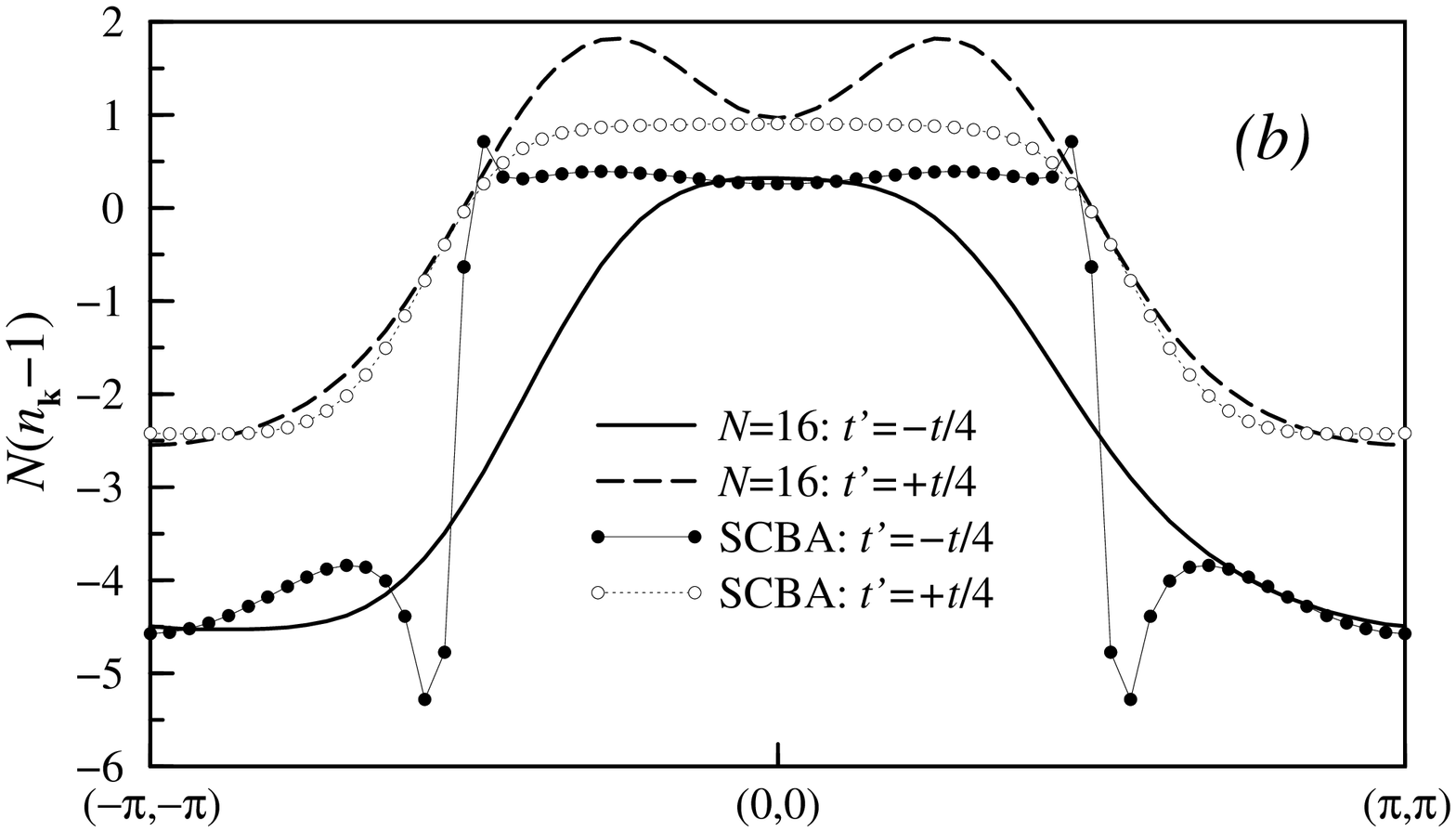}
\vskip -2.5 cm    
\includegraphics[width=0.8\linewidth,angle=-90]{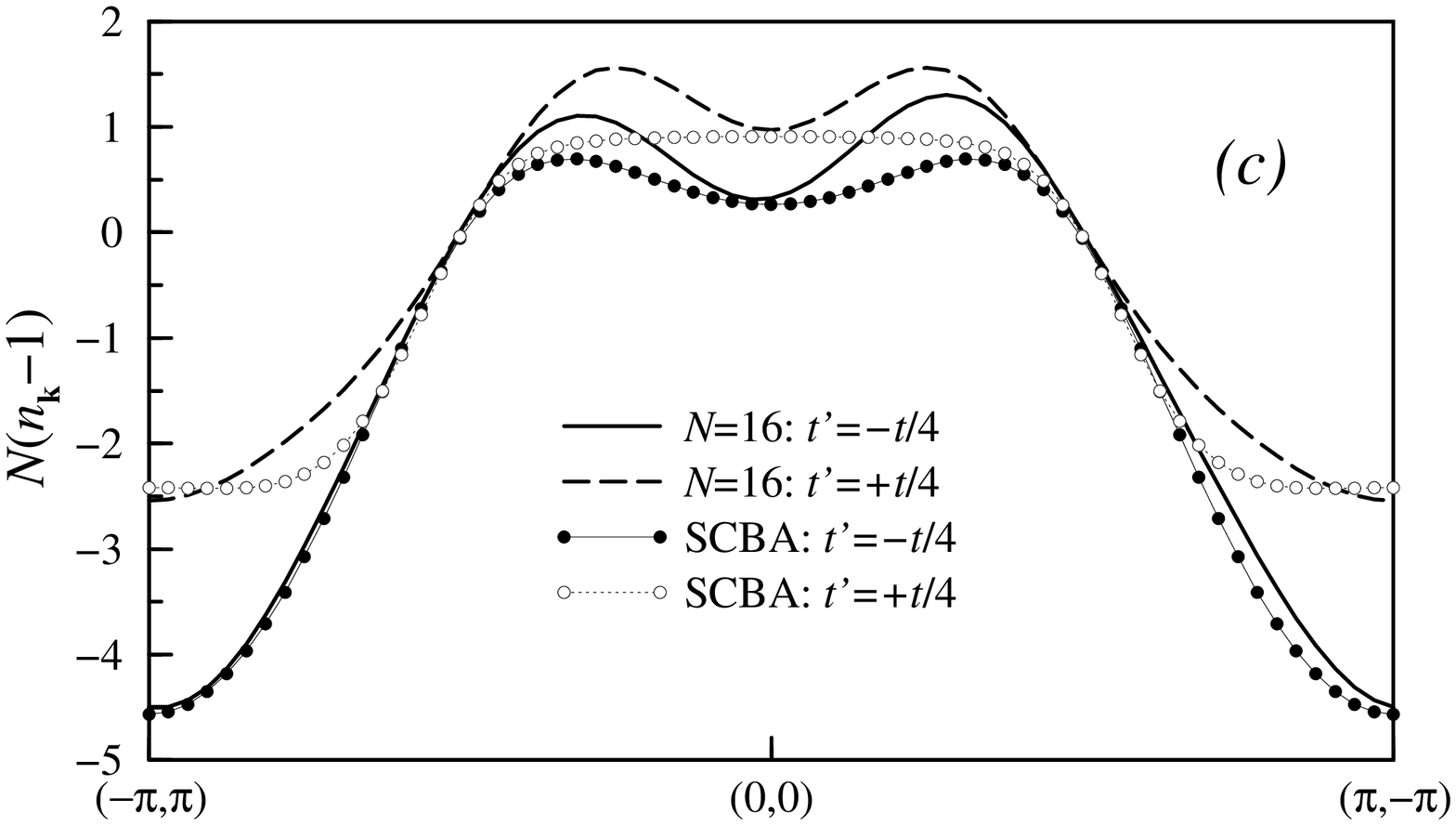}
\caption{ $N (n_{\bf k}-1)$ obtained from ED 
for various systems $N=16, 18, 20$ and $N=32$ from 
Ref.~[3]. For the SCBA $N\leq64\times64$,
$\gamma=0.999$ and note that delta-function contributions 
at ${\bf k}=\pm {\bf
k}_0$ are not shown.
(a) $t$-$J$ model, $t'=0$,
(b) $t$-$t'$-$J$ model with $t'=\pm t/4$
for ${\bf k}\in[(-\pi,-\pi) \to (\pi,\pi)]$,
and (c) for ${\bf k}\in[(-\pi,\pi)
\to (\pi,-\pi)]$.
}
\end{center}\end{figure}
\end{document}